\documentclass[a4paper,superscriptaddress,twocolumn,prb,leqno,showkeys]{revtex4-1}

\usepackage{float}
\usepackage{amsmath}
\usepackage{graphicx}
\usepackage{color}
\usepackage{epstopdf}

\begin{document}
\title{Signatures of phase transitions in the microwave response of YbRh$_2$Si$_2$}

\author{Katja Parkkinen}
\affiliation{1.\ Physikalisches Institut, Universit\"at Stuttgart, Stuttgart, Germany}
\affiliation{University of Helsinki, Helsinki, Finland}
\author{Martin Dressel}
\affiliation{1.\ Physikalisches Institut, Universit\"at Stuttgart, Stuttgart, Germany}
\author{Kristin Kliemt}
\affiliation{Physikalisches Institut, Goethe-Universit\"at Frankfurt, Frankfurt/Main, Germany}
\author{Cornelius Krellner}
\affiliation{Physikalisches Institut, Goethe-Universit\"at Frankfurt, Frankfurt/Main, Germany}
\affiliation{Max Planck Institute for Chemical Physics of Solids, Dresden, Germany}
\author{Christoph Geibel}
\author{Frank Steglich}
\affiliation{Max Planck Institute for Chemical Physics of Solids, Dresden, Germany}
\author{Marc Scheffler}
\affiliation{1.\ Physikalisches Institut, Universit\"at Stuttgart, Stuttgart, Germany}
\email{scheffl@pi1.physik.uni-stuttgart.de}

\begin{abstract}
We used a spectroscopic microwave technique utilizing superconducting stripline resonators at frequencies between 3~GHz and 15~GHz to examine the charge dynamics of YbRh$_2$Si$_2$ at temperatures and magnetic fields close to the quantum critical point. The different electronic phases of this heavy-fermion compound, in particular the antiferromagnetic, Fermi-liquid, and non-Fermi-liquid regimes, were probed with temperature-dependent microwave measurements between 40~mK and 600~mK at a set of different magnetic fields up to 140~mT. Signatures of phase transitions were observed, which give information about the dynamic response of this peculiar material that exhibits field-tuned quantum criticality and pronounced deviations from Fermi-liquid theory.
\end{abstract}


\maketitle
\section{Introduction}
\label{sect:introduction}

Quantum phase transitions (QPTs) and quantum criticality are major topics of present research in solid state physics.
Several heavy-fermion metals have been studied in great detail in this context and are now considered model systems where magnetic QPTs lead to unconventional behavior of the conduction electrons, which is generally named non-Fermi-liquid (NFL) behavior \cite{degiorgi1999electrodynamic,loehneysen2007fermi}.
Amongst these materials, YbRh$_2$Si$_2$ has attracted substantial attention: it is a stoichiometric, tetragonal compound that orders antiferromagnetically at $T_N$ = 70~mK in the absence of a magnetic field, and this antiferromagnetic (AFM) order can be suppressed smoothly with application of an external magnetic field; if the field is applied in the tetragonal plane, a very small field of 60~mT suffices to suppress $T_N$ down to zero at the quantum critical point (QCP).
At higher field, well-defined Fermi-liquid (FL) behavior is observed, e.g.\ a clear $T^2$ temperature dependence of the dc resistivity. At temperatures above the AFM and FL phases, pronounced NFL behavior is observed, for example a linear dependence of the dc resistivity as a function of temperature was found covering up to three orders of magnitude in temperature. In the middle of the NFL regime, an additional temperature scale $T^*$ was identified and interpreted as a transition from a small Fermi surface to a large Fermi surface upon increasing field \cite{gegenwart2007multiple, paschen2004hall}, but this interpretation remains debated \cite{kummer2015temperature}.
The rich phase diagram of YbRh$_2$Si$_2$ at temperatures well below 1~K has been studied with numerous thermodynamic, magnetic, and transport measurements, but this particular regime of finite, small magnetic fields and very low temperature was so far inaccessible to spectroscopic techniques. This is in contrast to higher temperatures, where scanning tunneling spectroscopy \cite{ernst2011emerging}, neutron spectroscopy \cite{stock2012incommensurate}, and angle-resolved photoemission spectroscopy \cite{kummer2015temperature} have lead to important results concerning the electronic and magnetic properties of YbRh$_2$Si$_2$.
NMR spectroscopy can reach temperatures well below 100~mK, but for YbRh$_2$Si$_2$ so far has been limited to magnetic fields higher than the critical field \cite{ishida2002b, kambe2014degenerate}. ESR spectroscopy, which has yielded important information about the spin dynamics in YbRh$_2$Si$_2$ at temperatures above 0.5~K \cite{krellner2008relevance, sichelschmidt2003low, sichelschmidt2010low}, only recently approaches the phase space close to the QCP \cite{scheffler2013microwave}.

\begin{figure*}[tbp]
	\begin{centering}
	\includegraphics[width=0.8\textwidth]{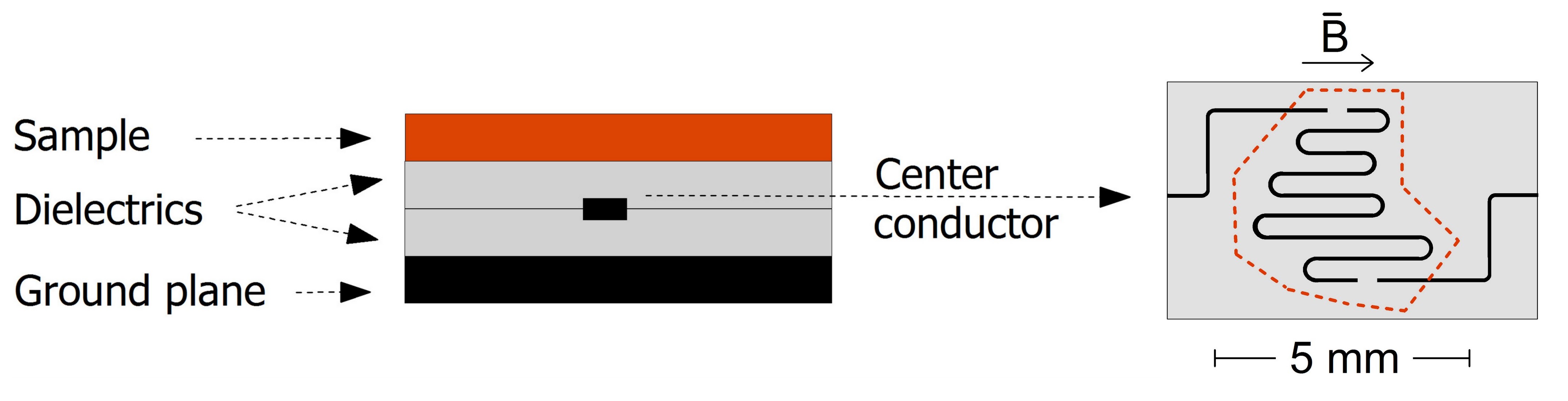}
	\caption{Scheme of stripline microwave resonator: cross section of a stripline (left) and a top view of the resonator plane with the shape of the YbRh$_2$Si$_2$ sample indicated by the dashed line (right).}
	\label{fig:stripline}
	\end{centering}
\end{figure*}

Here we now demonstrate how spectroscopic microwave experiments concerning the charge dynamics, i.e.\ the electrical conduction at GHz frequencies, can be performed on YbRh$_2$Si$_2$ at temperatures and fields close to the QCP. Microwave spectroscopy on heavy fermions addresses the fundamental electrodynamic properties of these materials \cite{scheffler2013microwave,basov2011electrodynamics}: the transport relaxation rate of the conduction electrons, manifest as the Drude response in the frequency-dependent conductivity \cite{dressel2006verifying}, is reduced by orders of magnitude compared to conventional metals \cite{degiorgi1999electrodynamic, millis1987large} and for certain heavy-fermion metals has been found to be as low a few GHz \cite{schefflerphysicaB2005,  scheffler2005extremely, schefflerphysicaB2006,  scheffler2010eurphys}. The electrodynamics of YbRh$_2$Si$_2$ has been studied previously by infrared spectroscopy \cite{kimura2006optical}, where indications for such a slow Drude relaxation were found, but it could not be directly observed. The very high effective mass of YbRh$_2$Si$_2$ indeed suggests an extremely low Drude relaxation rate \cite{scheffler2013microwave} at low temperatures, but on the other hand quantum critical fluctuations might cause a higher scattering rate. Furthermore, it is of general interest how the fundamentally different charge properties in YbRh$_2$Si$_2$ in the AFM, NFL, and FL phases are reflected in the dynamic response. For example, for a FL it is expected that the relaxation rate depends quadratically on both temperature and frequency \cite{degiorgi1999electrodynamic, dressel2002electrodynamics, gurzhi1959mutual, rosch2005zero}, while the NFL regime with its peculiar linear temperature dependence of the relaxation rate might also feature an unconventional frequency dependence. In this case, a transition between FL and NFL regimes at GHz frequencies might have different signatures than those studied previously in great detail for dc transport\cite{gegenwart2002magnetic, gegenwart2008unconventional, oeschler2008low}. In the present work, we therefore study the microwave response of YbRh$_2$Si$_2$ as a function of temperature at a set of different magnetic fields and thus probing the different electronic phases.

\section{Experiment}
The compound was studied with a spectroscopic method dedicated for GHz frequencies, utilizing a planar stripline resonator. A stripline is a microwave transmission line conceptually similar to a coaxial cable \cite{diiorio1988rf, hafner2014surface,  oates1991surface}, consisting of a strip-shaped inner conductor that is surrounded by two dielectric slabs and two outer conductor planes, as schematically seen in Figure \ref{fig:stripline}. 
In our measurement configuration, one of the ground planes is replaced by the conductive sample under study, in this case a single crystal of YbRh$_2$Si$_2$. Being part of the transmission line structure that carries the microwave signal, the sample affects the microwave field. Thus, the method is sensitive to the charge dynamics of the sample at GHz frequencies, which in the present case of a bulk metal become manifest to the measured observables via the skin effect \cite{dressel2002electrodynamics}. Since we require that the sample is the main contributor to the measured losses, the other conducting parts of the resonator are made of superconducting lead and the dielectric slabs are fabricated from low-loss \cite{krupka1999complex} sapphire wafers. 
In order to achieve the high sensitivity needed to study low-loss materials, the interaction between the sample and the electromagnetic field of the transmission line has to be enhanced. To that end we utilize a high-Q resonant structure by fabricating two coupling gaps (50~$\mu$m in this work) into the center strip, thereby composing a resonator that is capacitively coupled to the outer transmission lines and supports standing waves with resonance frequencies depending the resonator length $L$:

\begin{equation}
\nu_{0,n}=\frac{nc}{2\sqrt{\epsilon}L},
\end{equation}
where $n$ denotes the mode number, $c$ the speed of light in vacuum, and $\epsilon$ the permittivity of the dielectric. The center conductor of the resonator is meandered under an area that is defined by the size of the sample. This way, one obtains the largest possible resonator length and lowest base frequency, in the present work 3~GHz. The meander shape of the resonator was individually designed for the sample, and the lead resonator was thermally evaporated onto the lower sapphire slab using laser-cut shadow masks of steel. The stripline configuration with the sample is then assembled into a sample box of brass and connected to outer 50~$\Omega$ microwave lines. A corresponding characteristic impedance of the stripline was achieved by sapphire plane (12 x 10 mm\textsuperscript{2}) thickness of 127~$\mu$m, lead strip thickness of 1~$\mu$m, and stripline width of 45~$\mu$m.
With the static magnetic field applied in the plane of the resonator to mitigate the detrimental effect of the magnetic field on the performance of the superconducting resonator \cite{bothner2012magnetic}, one cannot exclude completely the excitation of ESR in the sample \cite{scheffler2013microwave,clauss2015optimization}. However, for the combinations of static magnetic field, microwave frequency, and temperature in this study we did not observe pronounced signs of ESR. Thus we expect that here, as usual for highly conductive metals, the charge response at GHz frequencies clearly dominates over the spin response.

The experiments were performed in a \textsuperscript{3}He\slash\textsuperscript{4}He dilution refrigerator with a microwave inset in a temperature range of 40~mK -- 600~mK and at magnetic fields of up to 140~mT. The microwave is generated and the signal transmitted through the resonator is measured by a vector network analyzer. From the transmission spectra, we determine the resonance frequency $\nu_0$ and the quality factor $Q$ of the resonator. In the desired case where the overall losses in the resonator are governed by the sample and all other loss contributions can be neglected, the surface resistance $R_s$ of the sample is related to the two measured quantities as
\begin{equation}
R_s=G \frac{\nu_0}{Q}, \label{eq:Rs}
\end{equation}
where $G$ is a factor depending the geometry of the stripline \cite{hafner2014surface}.

\section{Results}
A total of five resonator modes ranging from 3~GHz to 15~GHz were observed and measured at five fixed magnetic fields between 0~mT and 140~mT applied parallel to the tetragonal plane of the YbRh$_2$Si$_2$ crystal with the resonator along the same plane. Although lead is a type-I superconductor and has a critical magnetic field of 80~mT, the stripline maintained a resonance of a measurable quality factor at even higher fields.
Presently we do not know to which extent we can also assume for these magnetic fields above 80~mT that the total losses in the resonator are dominated by the sample and Eq.\ \ref{eq:Rs} still holds. However, since the temperature in the present study is below 1~K, while the zero-field critical temperature of lead is above 7~K, we can assume that any pronounced temperature dependence in the measured resonator response is caused by the Yb$_2$Rh$_2$Si$_2$ sample and not by the lead.

\begin{figure}[tbp]
	\begin{centering}
	\includegraphics[width=0.45\textwidth]{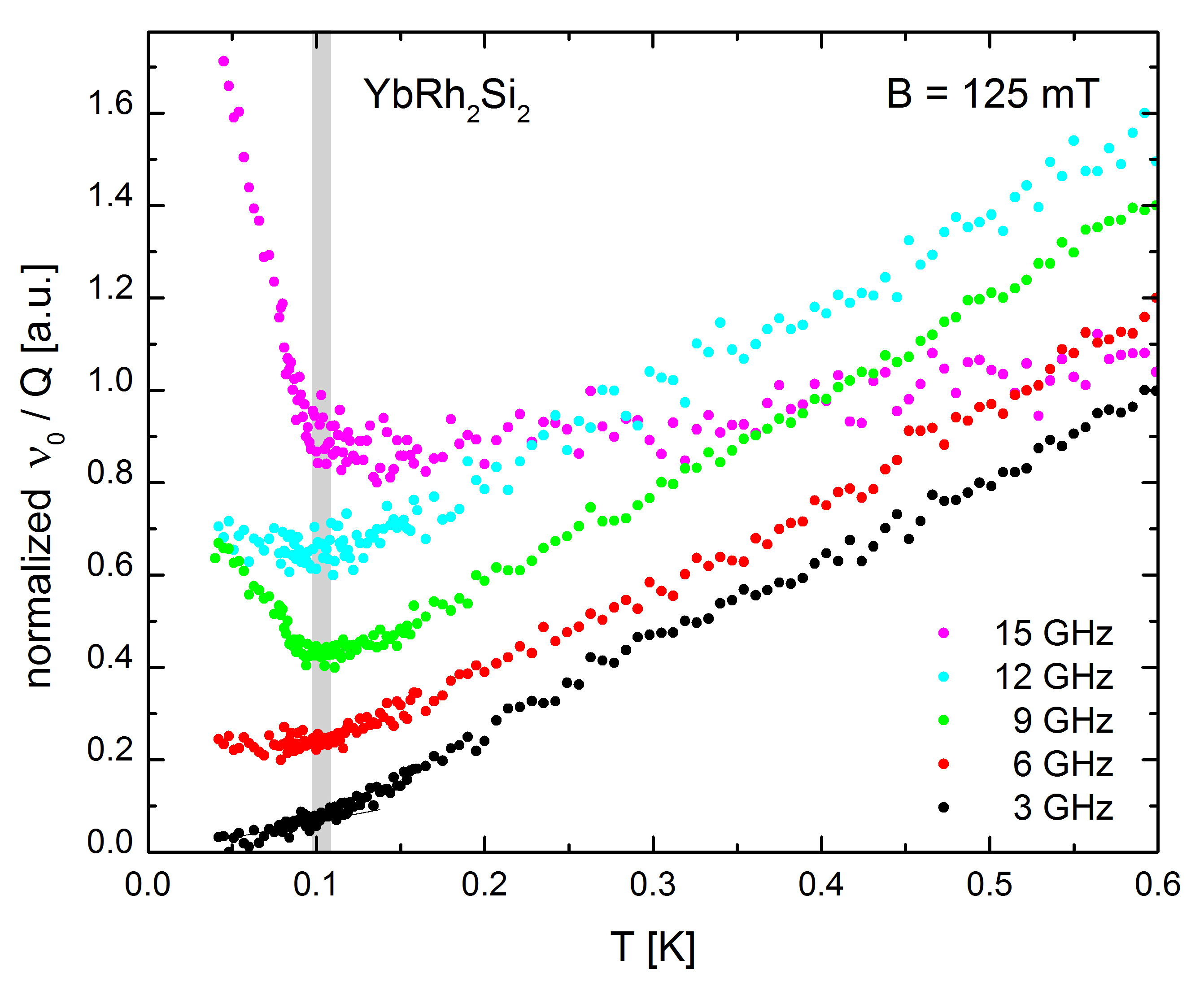}
	\caption{The transition between Fermi- and non-Fermi-liquid phases shows up as a clear change of trend in the ratio between resonance frequency and quality factor. The pictured transition is in a magnetic field of 125~mT. Data are normalized and shifted for clarity.}
	\label{fig:125mT}
	\end{centering}
\end{figure}

Fig.~\ref{fig:125mT} presents the evolution of the ratio between measured resonance frequency $\nu_0$ and quality factor $Q$ for different resonator modes in a magnetic field of 125~mT. The absolute magnitudes of these data were normalized and shifted as [$V$-$V_{min}$]/[$V_{max}$-$V_{min}$]+ $C$, where $V=\nu_0$\slash $Q$ and $C$ is a constant corresponding to the magnitude of the resonance frequency $C(\nu_0)=(\nu_0\slash$3~GHz~-1)~$\cdot$~0.2, for visual clarity. 
In the ideal case that Eq.\ \ref{eq:Rs} holds also at this field, the temperature dependence of $\nu_0 \slash Q$ is directly proportional to the temperature dependence of the surface resistance $R_s$.
If a transition from one electronic phase to another affects the electronic transport of YbRh$_2$Si$_2$, it should also show up in our microwave measurements. Unfortunately, establishing a quantitative relation from experimentally accessible surface resistance $R_s(\nu)$ to conductivity $\sigma(\nu)$ is quite challenging because it is not clear \textit{a priori} in which of the skin effect regimes that can occur in metals at low temperatures we operate \cite{dressel2002electrodynamics, hafner2014cecu6}.

The temperature-dependent data in Fig.\ \ref{fig:125mT} clearly indicate a change in electronic properties around 100~mK, which is the transition temperature between FL- and NFL-regimes. In particular, the dc resistivity features a smooth transition from low-temperature quadratic behavior to linear behavior at higher temperatures \cite{gegenwart2002magnetic,custers2003break}, and our 3~GHz data is rather reminiscent of such behavior. For higher frequencies, though not evolving completely consistently from one frequency to the next, there seems to be a trend towards more pronounced changes at this temperature of 100~mK. Considering that at finite frequencies the FL resistivity is expected as $\rho(T,\nu) = \rho_0 + a (4\pi^2 (k_B T)^2 + (h \nu)^2)$ \cite{scheffler2013microwave,gurzhi1959mutual} while for the NFL regime linear temperature dependence was determined for dc transport and a similar frequency dependence might be expected, such a stronger transition in the temperature dependence at higher frequencies might indeed signal finite-frequency characteristics of the FL-NFL transition. From this point of view, it could also be plausible if the experimental signatures of the FL-NFL transition occurred at slightly different temperatures for different probing frequencies, but we cannot conclude anything in this respect within our present experimental error bars.

\begin{figure*}[tbp]
    \centering
        \includegraphics[width=0.8\textwidth]{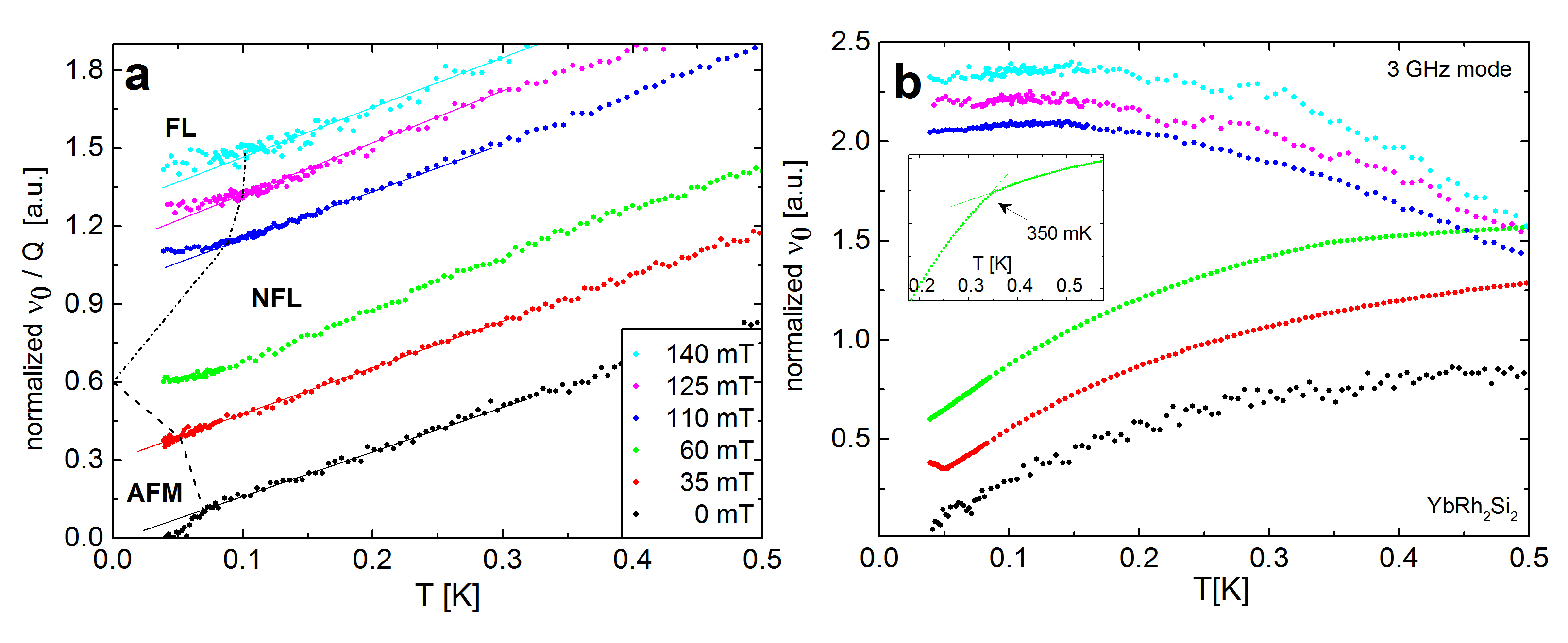}
\caption{3 GHz resonance mode in different magnetic fields: \textbf{a}.\ The ratio between resonance frequency and quality factor. Deviations from the linear trends are shown with the dotted curve, and can be associated with phase transitions. \textbf{b}.\ The evolution of the resonance frequency. The inset shows a change of trend in the measured resonance frequency at 60~mT and 350~mK. The absolute values were normalized and shifted for clarity. }
\label{fig:main}
\end{figure*}

In Fig.~\ref{fig:main} we plot the temperature dependence of $\nu_0 \slash Q$ and resonator frequency $\nu_0$ for a set of different magnetic fields for a single microwave frequency of 3~GHz, with normalization as [$V$ -$V_{min}$]/[$V_{max}$-$V_{min}$]+ $C$, where $C$ is a constant corresponding to the magnitude of the magnetic field $C(B)=B$[mT]$\slash$100~mT and $V$ the normalized physical quantity: $\nu_0$\slash $Q$ (left) and $\nu_0$ (right). Considering the behavior discussed for Fig.\ \ref{fig:125mT}, for this comparably low frequency one can expect behavior in $\nu_0 \slash Q$ that resembles the established dc resistivity behavior. Indeed, in the zero-field measurement in Fig.\ \ref{fig:main}a one can observe a clear kink in the temperature dependence around 70~mK that indicates the AFM transition. Also, the signature of the FL-NFL transition moves to higher temperatures as the magnetic field is increased from 110~mT to 140~mT, as expected. These observed phase changes agree with those previously measured with e.g.\ electric resistivity \cite{gegenwart2002magnetic} or ac-susceptibility experiments \cite{trovarelli2000ybrh}.
As visible from the straight lines that act as guides to the eye in Fig.~\ref{fig:main}a, $\nu_0 \slash Q$ exhibits quite linear temperature dependence over extended temperature ranges in the NFL regime. While such linear behavior is established for the dc resistivity $\rho_{dc}$, it comes as surprise for $\nu_0 \slash Q$ because $R_s$ exhibits a weaker temperature dependence than $\rho_{dc}$ in both the normal and anomalous skin effect regime, at least for conventional metals \cite{dressel2002electrodynamics, pippard1947surface}.

While Fig.\ \ref{fig:main}a represents the surface resistance, the resonator frequency plotted in Fig.\ \ref{fig:main}b depends on the skin depth of the sample: longer skin depth means further microwave penetration into the sample, corresponding to larger mode volume of the resonator, which in turn means lower resonator frequency $\nu_0$. In general, metals conduct better at lower temperatures and correspondingly one expects an increase of the resonator frequency upon cooling, as is commonly observed for metals and superconductor \cite{hafner2014surface, hafner2014cecu6, pozar2009microwave, ma1993microwave}.
Surprisingly, we find such behavior in $\nu_0$ only for magnetic fields above 100~mT, larger than the critical field of lead, whereas for fields of 60~mT and lower, we find the opposite trend.

As a side-note, in addition to the examined transitions between AFM, NFL, and FL phases, a feature around 350~mK in magnetic field of 60~mT appears as a clear change of trend in the resonance frequency, as is shown in the inset of Fig.~\ref{fig:main}b.
A similar phenomenon has been reported in dc transport at this temperature range \cite{reid2014wiedemann}: In 60~mT, below a characteristic temperature of 350~mK both the electric and thermal resistivities were shown to deviate from the linear trend associated with a non-Fermi liquid behavior. 
Anomalies in this temperature range were also observed previously in the heat capacity \cite{oeschler2008low,custers2003break}.

\section{Conclusions and Outlook}
We used a spectroscopic microwave technique utilizing superconducting stripline resonators to examine the charge dynamics of YbRh$_2$Si$_2$ at GHz frequencies and at temperatures and magnetic fields close to the quantum critical point. The different electronic phases were probed with a set of different magnetic fields, and signatures of transitions between AFM, NFL, and FL phases were clearly observed in the microwave response. 

The described method offers a tool for examining the dynamic response of the material exhibiting quantum criticality and pronounced deviations from the Fermi liquid theory. In particular, the photon energy of the employed microwave radiation corresponds to the other energy scales of this experiment, namely temperature and magnetic field, and therefore addresses the intrinsic physics that govern the material properties of YbRh$_2$Si$_2$. Future experimental work should study the microwave response at fixed temperatures as a function of magnetic field, which will require detailed understanding of the field-dependent behavior of the superconducting lead stripline resonator \cite{scheffler2013microwave}.

\section*{Acknowledgements}
We thank G. Untereiner for resonator fabrication and mounting. This work was funded by the DFG.

\normalsize
\label{sect:bib}
\bibliographystyle{unsrt}
\bibliography{bibfile_2015-09-08}

\end{document}